\documentclass[aps,footinbib,twocolumn,showpacs,amsmath,amssymb,superscriptaddress,prb,groupedaddress]{revtex4}
\usepackage{graphicx,amsmath}
\usepackage{epstopdf}
\usepackage{amssymb}
\usepackage{grffile}
\usepackage[usenames]{color}
\usepackage{indentfirst}
\usepackage{float}
\usepackage{color}
\usepackage{mathrsfs}
\usepackage{dcolumn}
\usepackage{bm}
\usepackage[colorlinks=true,bookmarks=false,citecolor=blue,linkcolor=red,urlcolor=blue]{hyperref}

\begin{document}
\title{The fractional quantum Hall nematics on the first Landau level in a tilted field}
\author{Dan Ye}
\author{Chen-Xin Jiang}
\author{Zi-Xiang Hu}
\email{zxhu@cqu.edu.cn}
\affiliation{Department of Physics and Chongqing Key Laboratory for Strongly Coupled Physics, Chongqing University, Chongqing 401331, People's Republic of China}
\begin{abstract}
 We investigated the behavior of fractional quantum Hall (FQH) states in a two-dimensional electron system with layer thickness and an in-plane magnetic field.  Our comparisons across various filling factors within the first Landau level revealed a crucial observation. A slight in-plane magnetic field specifically enhances the nematic order of the $\nu = 7/3$ FQH state.  For this particular filling, through calculating the energy gap, the Ising nematic order parameter, the pair-correlation function, and the static structure factor, we observed that as the in-plane magnetic field increases, the system first enters into an anisotropic FQH phase without closing the spectrum gap, then the FQH nematic (FQHN) phase after neutral gap closing.  The system eventually enters a gapless one-dimensional charge density wave (CDW) phase for a large in-plane field. We thus provide a full phase diagram of the $\nu = 7/3$ state in a tilted magnetic field, demonstrating the existence of the FQHN, which aligns with recent resonant inelastic light scattering (RILS) experimental observations.

\end{abstract}
\date{\today}
\maketitle

 \section{Introduction}
The fractional quantum Hall effect (FQHE) remains one of the most interesting strongly correlated systems for electrons moving in an effective two-dimensional manifold.~\cite{D. C. Tsui,R. Prange} Generally speaking, the ground states of the FQH liquid are incompressible with topological order, that is, order without dependence on any symmetries.~\cite{R. B. Laughlin83} Additionally, spontaneous breaking of symmetry within the FQH liquid can potentially give rise to electronic liquid crystal phases, such as nematic, smectic, and stripe order.~\cite{E. Fradkin99,E. Fradkin00,L. Radzihovsky02,K. Sun08, D. G. Barci02320,M. J. Lawler04,A. M. Ettouhami06,K. Tsuda07,Q. Qian17} A particular area of interest arises when these distinct phases coexist, leading to quantum liquid crystal phases like fractional quantum Hall nematic phases. The concept of fractional quantum Hall nematic effect (FQHN) was originally proposed by Balents~\cite{L. Balents96} which was recently discovered in experiments.~\cite{J. Xia11, X. Fu20}  These phases have been experimentally observed in transport measurements.~\cite{M. P. Lilly99} By manipulating factors such as the in-plane magnetic field,~\cite{J. Xia10,Y. Liu13} valley occupancy through the application of in-plane strain~\cite{M. S. Hossain23,M. S. Hossain18,M. Shayegan06} or magnetoresistance measurements under hydrostatic pressure~\cite{N.Samkharadze15,K. A. Schreiber17,K. A. Schreiber18} researchers have observed systems exhibiting anisotropic longitudinal resistivity that is enhanced at low temperatures, along with a robust Hall conductivity plateau. The nematic state is fully translationally invariant but lacks rotational invariance. It can be visualized as stripes that fluctuate strongly and are riddled with dislocation defects but retain a preferential alignment in one direction. To definitively confirm the existence of a nematic phase, it is crucial to simultaneously investigate both rotational symmetry and translational invariance. However, previous studies have mainly focused on the breaking of rotational symmetry without achieving long-range translational invariance. Only recently, with the work reported in Ref.~\onlinecite{L. Du19}, there has been simultaneous evidence of broken rotational symmetry and translational invariance, allowing the identification of nematic phases.

Theoretical investigations, including the Hartree-Fock approximation and effective field theories, have predicted a range of phases in addition to the expected Wigner crystals. These include stripe phases, bubble phases, and other crystalline states.~\cite{A. A. Koulakov96, M. M. Fogler96, A. H. MacDonald00, M. M. Fogler02, D. G. Barci02,S.-Y. Lee02} These predictions have been further elaborated upon by various field theory approaches, which describe the incompressible nematic phase using an effective gauge theory~\cite{Mulligan10, Mulligan11} or by assuming the softening of the magnetoroton mode.~\cite{J. Maciejko13, Y. You14} These theories aim to capture the topological order and the nematic order resulting from spontaneous symmetry breaking. Microscopic theories, on the other hand, focus on analyzing the microscopic properties of FQHN in the thermodynamic limit, providing the necessary conditions for the robustness of the nematic fractional quantum Hall effect in microscopic Hamiltonians.~\cite{Bo2020} Recent numerical studies have further supported the existence of FQHN phases in both bosonic and fermionic systems, with the tuning of pseudopotential coefficients in a model Hamiltonian.~\cite{N. Regnault17, T. GraA18, Songyang}

The FQH states within the first Landau level (1LL) exhibit smaller energy gaps and are generally more prone to perturbations compared to those found in the lowest Landau level (LLL), as documented in various studies.~\cite{Wang12, Yang12, Rezayi00, Faugno21, Yang18} Consequently, the spontaneous emergence of nematic order, which involves the spontaneous breaking of rotational symmetry without a preferred direction, may be more feasible in higher LLs due to the reduced energetic barrier required for its stabilization.  Fractional quantum Hall (FQH) states exhibit inconsistent responses depending on the filling factor. In half-filling $\nu = 5/2$, the ground state is described as a p-wave paired superfluid, as supported by numerous studies.~\cite{G. Moore91, R. H. Morf98, A. Stern10, C. Nayak08} However, in the presence of a sufficiently large in-plane magnetic field, the 5/2 FQH state transitions into a compressible nematic phase.~\cite{M. P. Lilly99, W. Pan99, Eisenstein88, Rezayi00, Friess14} On the other hand, at $\nu = 7/3$, the application of even small in-plane magnetic fields leads to a pronounced transport anisotropy that coexists with the quantized Hall plateau.~\cite{J. Xia11} To investigate the FQH nematic phase at various filling factors within the first Landau level (1LL), we chose to utilize the torus geometry with translational invariance. The introduction of an in-plane magnetic field to the electrons introduces anisotropy, thereby disrupting the rotational symmetry of the system. Recently, one of us generalized the description of pseudopotentials to incorporate non-zero non-diagonal pseudopotentials $c_{m,n}$ (where $m \neq n$)  that do not conserve angular momentum. This generalization enables the representation of system anisotropy in the absence of rotational invariance.~\cite{Haldane11, Yang17}

In this paper, we present a microscopic study in a realistic model examining the influence of in-plane magnetic field and layer thickness on FQH states within the 1LL, across different filling factors, especially at $\nu = 7/3$. The analysis is grounded in the pseudopotential description of electron-electron interaction within an in-plane magnetic field, accounting for finite layer thickness. The paper is organized as follows. Section II reviews the single-electron solution, delves into the pseudopotentials, and analyzes their behavior with varying parameters. Section III provides evidence for the existence of FQH states by calculating the energy spectrum at the filling factor $7/3$, evaluating the magnitude of the correlation function, and exploring the nematic phase within a specific parameter range. We obtain the nematic phase for different filling factors, findings that align with experimental results. Finally, Section IV concludes the article with a summary of our findings.

 \section{Model and Method}
  We explore the impacts of a magnetic field that possesses both a component perpendicular to the plane ($B_\perp$) and a component within the plane ($B_\parallel$). Without loss of generality, we assume that the in-plane field $B_\parallel$ is oriented along the $x$ axis. In particular, the in-plane magnetic field $B_\parallel$ influences the system only when the thickness of the two-dimensional electron gases (2DEG) is significant. For simplicity, we consider the electrons to be confined in the z-direction by a parabolic quantum well potential represented by $\frac{1}{2}m\tilde{\omega}_0z^2$. The thickness of the system is related to the characteristic width of the harmonic well, defined as $w_0=1/\sqrt{m \tilde{\omega}_0}$, in which $m$ denotes the effective mass of the electrons and $\tilde{\omega}_0$ is the angular frequency associated with the confinement potential in the $z$ direction. This assumption allows us to simplify the problem by focusing solely on the electron motion within the plane, while incorporating the finite thickness of the system through the parameter $w_0$. The Hamiltonian for a single particle is then expressed as
\begin{eqnarray}\label{h1}
H&=&\frac{1}{2m} \sum_{i=x,y,z} \left(p_i+eA_i\right)^2 +\frac{1}{2}m\tilde{\omega}_0^2z^2
\end{eqnarray}
Similar to the special case where $B_\parallel$ is absent, the Hamiltonian can be expressed in a diagonal form as
\begin{eqnarray}
H= \omega_1 X^\dagger X + \omega_2 Y^\dagger Y + \text{constant}.
\end{eqnarray}
where $(X, X^\dagger)$ and $(Y, Y^\dagger)$ are two sets of decoupled bosonic operators obeying the canonical commutation relation $[X, X^\dagger]=[Y, Y^\dagger]=1$ and $[X, Y]=0$. They are linear combinations of the canonical momentums $\pi_i = p_i + eA_i$ and $\pi_4 = m\tilde{\omega}_0 z$. The details of the diagonalization procedure can be found in Ref.~\onlinecite{yang17prb}.  The quasi-particle eigenenergy is given by $\omega_1^2 = \frac{1}{2} (\tilde{\omega}_0^2 + \omega_x^2 + \omega_z^2 - \sqrt{-4 \tilde{\omega}_0^2 \omega_z^2 + (\tilde{\omega}_0^2 + \omega_x^2 + \omega_z^2)^2})$, $\omega_2^2=\frac{1}{2}(\tilde{\omega}_0^2 + \omega_x^2 + \omega_z^2 +\sqrt{-4 \tilde{\omega}_0^2 \omega_z^2 + (\tilde{\omega}_0^2 + \omega_x^2 + \omega_z^2)^2})$. Here, $\omega_x$ and $\omega_z$ represent the cyclotron frequencies corresponding to $eB_\parallel/m$ and $eB_\perp/m$, respectively.  As a result, the single particle Landau wave functions are now indexed by two quantum numbers $\left | m,n  \right \rangle =\frac{1}{\sqrt{m!n!}}(X^{\dagger })^m(Y^{\dagger })^n\left | 0 \right \rangle$, where $|0\rangle$ is the vacuum state. In the limit of $\omega_x \rightarrow 0$, when $\tilde{\omega}_0 > \omega_z$, the operator $(X, X^\dagger)$ raises and lowers the in-plane Landau levels (LLs), while the operator $(Y, Y^\dagger)$ raises and lowers the harmonic modes along the $z$ axis (or the subbands). The roles of $X$ and $Y$ are reversed for $\tilde{\omega}_0 < \omega_z$.
\begin{figure}
 \includegraphics[width=8cm]{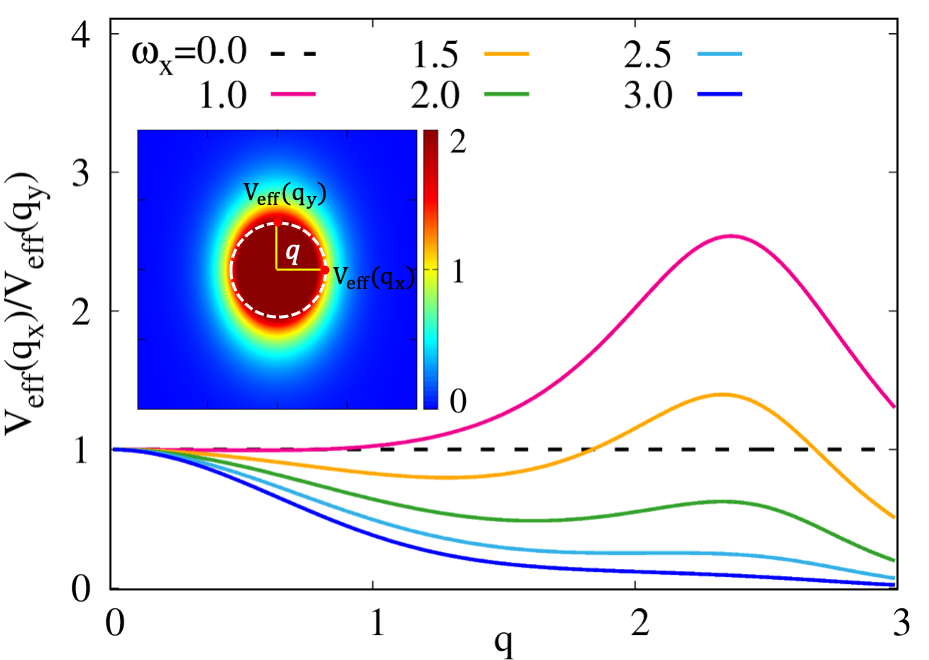}
 \caption{\label{vq}(color online)  When $w_0=1.25$, the ratio of effective interaction in two directions as a function of the circular radius $q$ under different parallel magnetic fields. The inset is the 2D distribution of $V_{\text{eff}}(\vec q)$ at $\omega_x=3.0$.  }
\end{figure}

The three-dimensional Coulomb interaction has the Fourier form $V_{\vec k}=1/k^2=1/(|\vec{q}|^2+q_3^2)$ where $\vec{q}$ is the two-dimensional vector. Its LLL projected Hamiltonian is
 \begin{eqnarray}\label{hint}
H_C=\int d^3kV_{\vec k}|F_{mn}(\vec q,q_3)|^2 \hat{\rho}_{\bf q}\hat{\rho}_{-{\bf q}}e^{-q^2/2}
\end{eqnarray}
 where $\hat{\rho}_{\bf q} =\sum_ie^{i\vec q\cdot {\bf r}_i}$ is the guiding center density and  $F_{m,n}(\vec q,q_3)$ is the form factor from the LLL projection.  The lowest LL is defined as $m = n = 0$ and the 1LL is for $m = 0, n=1$ or $m=1, n=0$. After integrating the $z$ component $q_3$, we thus obtain the effective two-dimensional interaction.~\cite{yang17prb,Yang18}
 \begin{eqnarray}\label{finalf}
V_{\text{eff}}(\vec q)=\int_{-\infty}^{\infty}dq_3\frac{1}{|\vec{q}|^2+q_3^2}|F_{mn}(\vec q,q_3)|^2\label{effective2d}
\end{eqnarray}
 For a two-body interaction without rotational symmetry,
it was found~\cite{Yang17} that a generalized pseudopotential description can be used to describe it.
\begin{eqnarray} \label{PPS}
 V^+_{m,n}(\mathbf{k}) &=& \lambda_n \mathcal{N}_{mn}(L_m^n(|k|^2) e^{-|k|^2/2} {\mathbf{k}}^n + c.c) \nonumber \\
 V^-_{m,n}(\mathbf{k}) &=& -i \mathcal{N}_{mn}(L_m^n(|k|^2) e^{-|k|^2/2} {\mathbf{k}}^n - c.c)
\end{eqnarray}
where $\mathcal{N}_{mn}$ is the normalization factors. The effective two-body interaction including the anisotropic one can be expanded as $V_{\text{eff}}(\mathbf{k}) = \sum_{m,n,\sigma}^\infty c^{\sigma}_{m,n} V^{\sigma}_{m,n}(\mathbf{k})$ with coefficient $c^{\sigma}_{m,n} = \int d^2 k V_{\text{eff}}(\mathbf{k})  V^{\sigma}_{m,n}(\mathbf{k})$.

\begin{figure} 
 \includegraphics[width=0.38\textwidth]{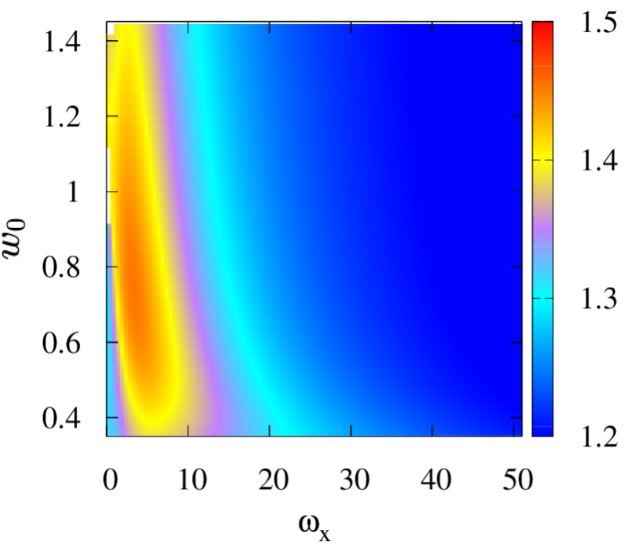}
 \caption{\label{pps}(color online) The ratio between the first two pseudopotential coefficients $c_1/c_3$  in 2D plot parameterized by $\omega_x$ and $w_0$. It exhibits a peak value at finite in-plane magnetic field. The value at the peak is approximately $c_1/c_3 \sim 1.5$, exceeding the Coulomb value of $1.32$, suggesting an increase in the gap at that point. }
\end{figure}

In Fig.~\ref{vq}, we display the ratio of the effective interactions along the $q_x$ and $q_y$ axes on a circle with a radius of $q$, considering various strengths of the parallel magnetic field with a fixed layer thickness of $w_0 = 1.25$. In particular, when $\omega_x$ is nonzero, the strength of the effective interaction exhibits anisotropy, with a ratio different from one. This anisotropy leads to a break of the rotational symmetry. Furthermore, as $\omega_x$ increases, the ratio changes from being greater than one to smaller than one, indicating a potential change in the orientation of the effective interaction anisotropy. Such anisotropies can be experimentally probed through measurements related to the ground-state structure factor and the neutral excitation gap, providing valuable insight into the physics of the system. Fig.~\ref{pps}(a) illustrates the pseudopotential coefficients $c_1/c_3$ on a two-dimensional graph while changing the values of $\omega_x$ and $w_0$. It is known that $c_1$ plays a crucial role in the Laughlin state which is the zero energy eigenstate of the model Hamiltonian with only $c_1 \neq 0$. We can see $c_1/c_3$ exhibits a peak value ($c_1/c_3 \sim 1.5$ comparing to the Coulomb value $1.32$) at a finite $\omega_x$, indicating that a finite in-plane magnetic field tends to stabilize the Laughlin state at $\nu = 7/3$. This phenomenon is also evident in the energy gap as shown below. It has previously been verified by experiment~\cite{Dean} in which a strong enhancement of the gap was observed in a wide 2D quantum well.    

\section{Numerical  results}
We consider a system with $N_e$ electrons in the torus geometry. The number of quantum fluxes $N_\phi$ fixes the filling factor $\nu = N_e / N_\phi$. The torus is spanned by the vectors $\hat{L}_x$ and $\hat{L}_y$ in two principle directions. Unless otherwise specified, we consider a rectangle torus with $90$ degrees angle between $\hat{L}_x$ and $\hat{L}_y$. The aspect ratio $L_x/L_y=1$ is set mainly at unity. For $N_e = pN$ particles in $N_\phi = qN$ orbits with a maximum common divisor $N$, the filling factor is $\nu = p/q$.  The many-body center-of-mass magnetic translation operator $\bar{T}(a) = \prod_{i=1}^{N_e} T_i(a)$ in which $T_i(a)$ is the magnetic translation operator for each electron. In Landau gauge, we have two good quantum numbers $t = \sum_{i=1}^{N_e} m_i$ mod $N_\phi$, the total momentum in the $y$ direction in units of $2\pi/L_y$, and $s$, the center of mass translational momentum in units of $2\pi/L_x$ which contributes the $q$-fold degeneracy in energy spectrum. That is, they obey the relation $\bar{T}(L_x/N) = e^{i2\pi s/N}$ and $\bar{T}(L_y/N) = e^{i2\pi t/N}$.
\subsection{Ising nematic order}
\begin{figure}[ht]
\includegraphics[width=0.5\textwidth]{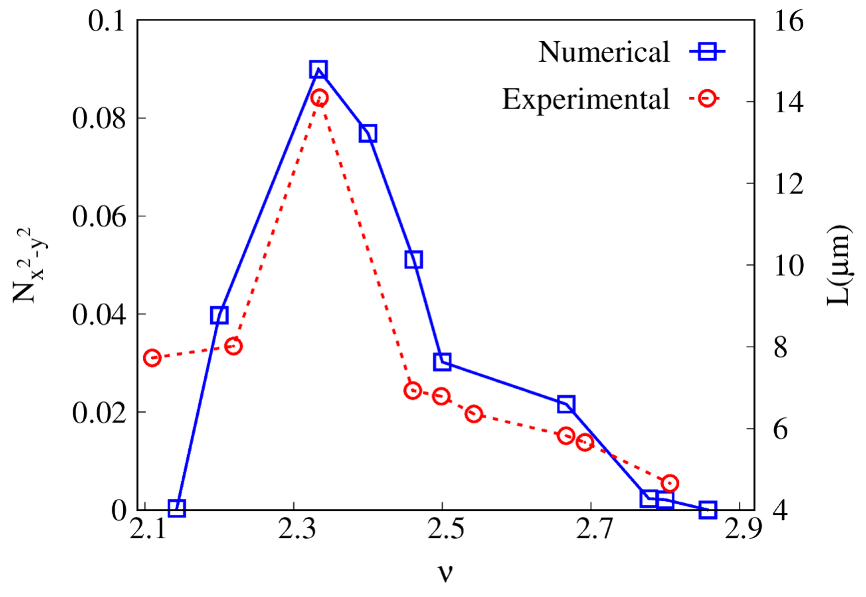}
\caption{\label{filling} The numerical calculation of the Ising nematic order parameters $N_{x^2-y^2}$ for $7-16$ electrons at various filling factors in the 1LL is represented by the empty blue squares. The experimentally observed plasmon coherence length $L$ in Ref.~\onlinecite{L. Du19} is represented by the empty red circles. }
\end{figure}

Once the FQH liquid enters the FQHN phase, the original full $SO(2)$ rotation symmetry is reduced to a discrete $C4$ rotation symmetry as a result of the periodic boundary conditions in the torus geometry. To quantitatively measure the nematic order after breaking the rotational symmetry, for a given ground state $|\psi\rangle$, we calculate the Ising nematic order parameter characterized by $d_{x^2-y^2}$ symmetry,~\cite{Abanin10, Metlitski10, N. Regnault17}
\begin{equation}
N_{x^2-y^2}=\frac{1}{N_e(N_e-1)}\sum_{\bf{q}}(\cos q_x-\cos q_y) \langle \psi | \bar{\rho}_{\bf{q}} \bar{\rho}_{-\bf{q}} | \psi  \rangle 
\end{equation}
where $\bar{\rho}_{\bf{q}}$ is the LLL projected guiding center density operator. This order parameter captures the anisotropy in the system, which is a hallmark of the nematic phase. By measuring this order parameter, we can gain insight into the nature of the nematic order and its evolution as a function of various parameters, such as the strength of the magnetic field or the layer thickness. In experiments, breaking the rotational symmetry leads to anisotropic transport properties. Recently, Du et al.~\cite{L. Du19} observed a pronounced plasmon intensity in the 1LL of the nematic phase, particularly at $\nu=7/3$, through measurements of long-wavelength spin-wave excitations using resonant inelastic light scattering (RILS) methods in a tilted magnetic field. The plasmon intensity is proportional to the square of the plasmon coherence length $L$. The significant value of $L$ at $\nu=7/3$ indicates the presence of long-range correlations that favor translational symmetry, providing important evidence for the FQHN. Meanwhile, a marked minimum intensity in the plasmon spectrum at $\nu = 5/2$ strongly suggests that the paired state overwhelms the competing nematic phases. Motivated by these findings, we conducted numerical investigations of the FQHN  in the 1LL, specifically targeting the filling fraction $\nu=7/3$ where the strongest nematic order is experimentally observed. Our aim is to gain a deeper understanding of the nematic phase and its properties in this regime, in the hope of providing further insights into the experimental observations and the underlying physics of the FQH nematic states.
 
In numerical calculations, for simplicity, we set the frequency in the $z$-direction, $\omega_z$ to unity. To obtain the ground state of the system, we employed exact diagonalization for tens of electrons. In Fig.~\ref{filling}, the Ising nematic order parameters for fixed tilted magnetic field and the layer thickness ($ \omega_x=26.5,w_0=1.25 $) at different filling factors in the 1LL are calculated. Here, we consider only the pure Coulomb interaction and neglect the influence of the effects from Landau-level mixing and disorder for simplicity. The number of electrons varies between $7$ and $16$, depending on the filling factor. For comparison, we also extracted experimental data from Ref.~\onlinecite{L. Du19} for the dependence of plasmon intensity on filling factor.  Our results qualitatively align with previous experimental findings, showing a peak in the filling factor of $7/3$, indicating the relative stability of FQHN in this filling. Therefore, our subsequent analysis will focus primarily on the $\nu = 7/3$ state.  

\subsection{Phase diagram}
\begin{figure}[ht]
\includegraphics[width=0.50\textwidth]{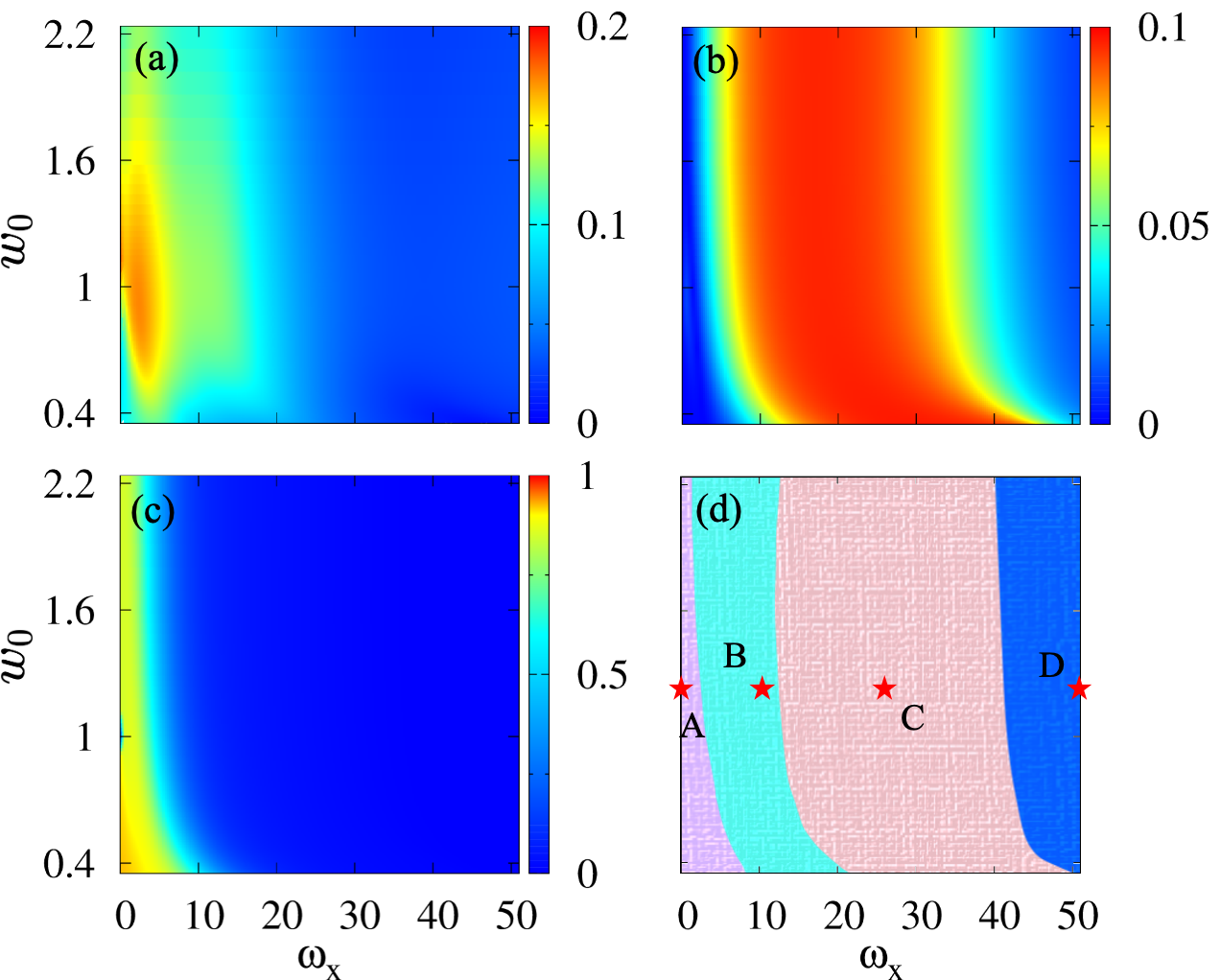}
\caption{\label{phase}(color online) Via scanning the parameters $\omega_x$ and $w_0$ for 10 electrons at $\nu = 7/3$, we calculate the neutral gap as the gap between the lowest two states in $k = 0$ subspace shown in (a). The nematic order parameters of the ground state are shown in (b). The ground state wave function overlap with that of the pure Coulomb interaction without in-plane field and thickness is shown in (c). (d) Schematic phase diagram labeled by A: isotropic FQH (IFQH), B: anisotropic FQH (AFQH), C: FQH nematics (FQHN) and D: charge density wave (CDW), respectively. }
\end{figure}

First, we discuss the Laughlin phase in which both the charge and neutral gaps are not closed. We simply define the neutral gap as the spectrum gap between the ground state and the first excited state in the $k=0$ subspacey. The results in the $\omega_x-w_0$ plane are illustrated in Fig.~\ref{phase}(a) for a system with 10 electrons. It shows that the gap is nonzero when $\omega_x$ is small and there is a maximum value for finite $\omega_x$, which is consistent with the pseudopotential $c_1/c_3$ as shown in Fig.~\ref{pps}(b). Even with a non-zero gap, the Laughlin phase can be separated into two components, one of which preserves rotational symmetry while the other does not. Therefore, the wave function overlap is computed between the ground state and the pure Coulomb ground state, which has no thickness and no magnetic field in the plane. As shown in Fig.~\ref{phase}(c), the isotropic Laughlin phase only survives for small $\omega_x$. This could be confirmed by examining the Ising nematic order $N_{x^2-y^2}$ of the ground state, as illustrated in Fig.~\ref{phase}(b), where the isotropic Laughlin phase, known as the A phase in Fig.~\ref{phase}(d), exhibits zero nematic order without breaking the rotation symmetry. 

\begin{figure}[ht]
\includegraphics[width=0.38\textwidth]{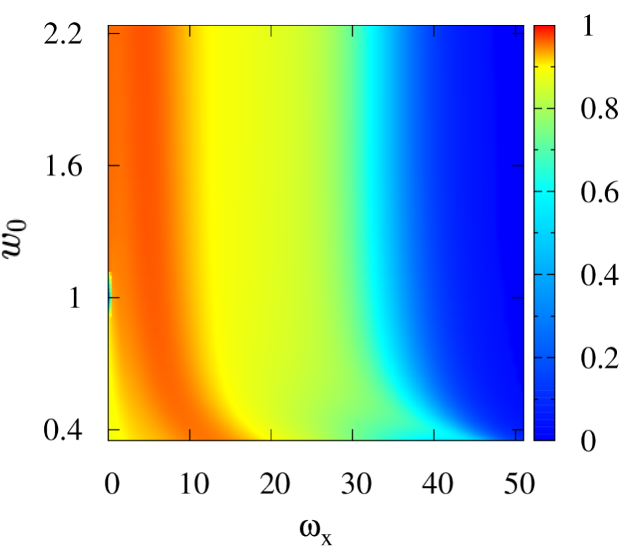}
\caption{\label{overlap}(color online) The maximum value of overlap between the $\nu  = 7/3$ ground state for each parameter point $(\omega_x, w_0)$, and the generalized Laughlin states parametrized by varying the metric $\gamma$. The generalized Laughlin wave function can be obtained by diagonalizing the pseudopotential Hamiltonian with $c_1(q_g)$. }
\end{figure}

The B phase, with a finite spectrum gap and nonzero nematic order, could be the anisotropic Laughlin phase. According to the geometric explanation provided by Haldane,~\cite{Haldane11} a set of anisotropic Laughlin states can be characterized by an intrinsic metric. These states, denoted as $\Psi_L^{\nu=1/m}(g) = \prod_{i < j} [b_i^\dagger(g)-b_j^\dagger(g)]^m \left | 0  \right \rangle $ where $b^\dagger(g)$ is the single particle creation operator in the anisotropic Landau basis,~\cite{QiuPhysRevB} represent the unique ground state with zero energy of the model Hamiltonian with nonzero $c_1$. Here, $g$ represents a unimodular metric associated with an ``area preserving" deformation.
 \begin{eqnarray}
 g=\left(
 \begin{array}{cc}
             \cosh2\theta + \sinh2\theta\cos2\phi & \sinh2\theta\sin2\phi \\
             \sinh2\theta\sin2\phi &  \cosh2\theta - \sinh2\theta\cos2\phi
            \end{array} \right),\nonumber
\end{eqnarray} 
where $\phi$ and $\theta$ represent the rotation and stretching parameters, respectively, for the primary motion of the electron in a magnetic field. By incorporating the $g$ metric, a circular trajectory transforms into an elliptical one.  In our case, we restrict the in-plane field in $x$ direction, thus the parameter $\phi$ should be a constant.  Consequently, we introduce a unified parameter $\gamma = \cosh2\theta + \sinh2\theta$ to characterize the metric, as discussed in the investigation of band mass anisotropy.~\cite{Yang12} The value of $\gamma = 1$ represents the isotropic scenario with rotational symmetry. To determine the intrinsic metric $\gamma_c$ of a given ground state wave function with fixed $\omega_x$ and $w_0$, it is necessary to compute the overlap $\mathcal{O}_g = \mathcal{O}(\phi, \gamma) = |\langle \Psi |\Psi_L(g)\rangle|^2$, where the intrinsic metric $g_c$ is the metric that maximizes this overlap. Here we produce the general Lauglin state $\Psi_L(g)$ by pseudopotential  $c_1(\overrightarrow{q_g}) = 2L_1(q^2_g)-2$ where $q_g^2=g_{ab}q^aq^b$ and $L_1(x)$ is the Laguerre polynomial. In this case, the value of $g_c$ is consistently set to $\phi = 0$, meaning that we only adjust $\gamma$ within the interval of $[1, 2.5]$. Fig.~\ref{overlap} shows the graph of the maximum $\mathcal{O}(\phi, \gamma)$ for the system. The red-colored area may represent the anisotropic Laughlin phase, as optimizing the metric $g$ can enhance the overlap to the identity.
Hence, in the B phase, the ground state shows a gap and can be characterized as an anisotropic Laughlin wavefunction with non-zero Ising nematic order as well. Here we need to notice that the overlap for zero in-plane magnetic field with $\omega_x=0$ is not very high $|\langle \Psi |\Psi_L(g=1)\rangle|^2 \simeq 69\%$ which mean the isotropic ground state is not very accurately described by Laughlin state in the 1LL. By increasing the layer thickness, the overlap could be significantly enhanced to $\sim 96\%$, indicating that the finite layer thickness plays a role in stabilizing the Laughlin state in the 1LL. Similar phenomenon was also observed previously in Ref.~\onlinecite{Peterson}.

\begin{figure}
 \includegraphics[width=0.48\textwidth]{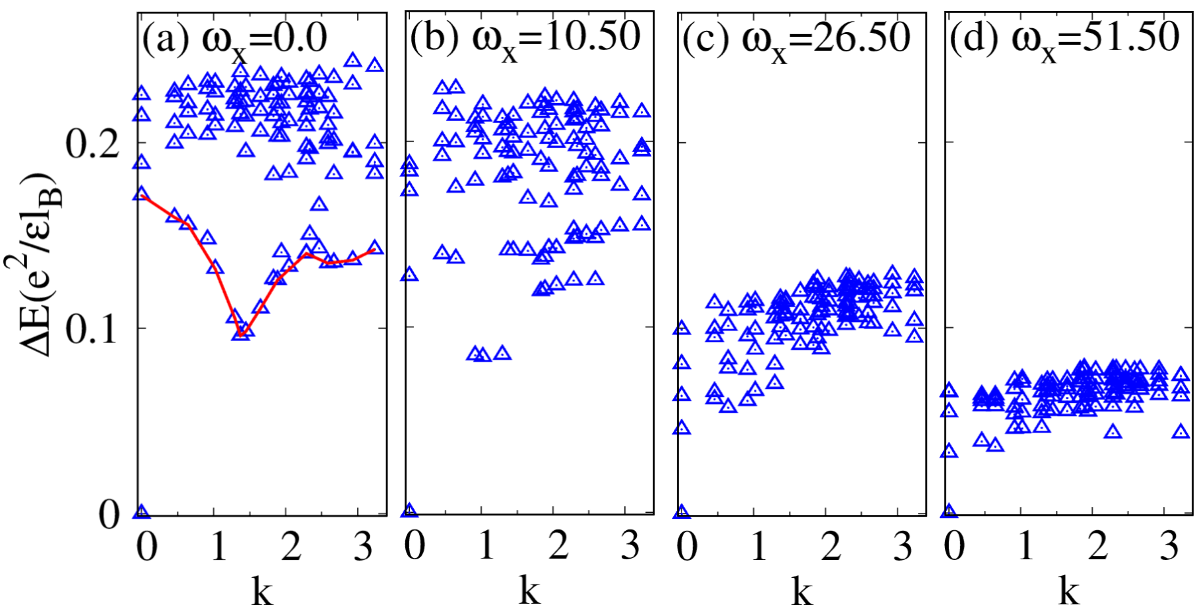}
\caption{\label{spectrum} The energy spectrum of 10 electrons for $\nu=7/3$ in the 1LL with fixed $w_0=1.25$ and different $\omega_x$. (a) $\omega_x=0.0$. (b) $\omega_x=10.5$. (c) $\omega_x=26.5$. (d) $\omega_x=51.5$. They are marked as asterisks in Fig.~\ref{phase}(d).  }
\end{figure}

Now we aim to demonstrate that the C phase in Fig.~\ref{phase}(d) corresponds to the FQHN phase. First of all, as shown in Fig.~\ref{phase}(b), it has a non-zero Ising nematic order. Fig.~\ref{phase}(a) shows a small but non-zero gap for a finite system in this region. This could be seen in the energy spectrum, as shown in Fig.~\ref{spectrum}. In which we fix $w_0 = 1.25$ and take different values of $\omega_x$. (a) $\omega_x=0.0$, (b) $\omega_x=10.5$, (c) $\omega_x=26.5$, and (d) $\omega_x=51.5$ represent four points (marked by asterisks in Fig.~\ref{phase}(d)) in phases A, B, C, and D respectively.  It is clear that the magnetoroton mode~\cite{A. Pinczuk93,M. Kang01,I. V. Kukushkin,S. M. Girvin85}becomes softer as the in-plane magnetic field increases, and the lowest excited state transitions to the $k = 0$ subspace in the C and D phase, although an energy gap still remains in a finite system.  To determine if the neutral gap has been eliminated, it is necessary to perform extrapolation in the thermodynamic limit.  In the following, we calculate the thermodynamic limit gap from analyzing the thermoelectric Hall conductivity scaling.

\begin{figure}[ht]
 \includegraphics[width=0.48\textwidth]{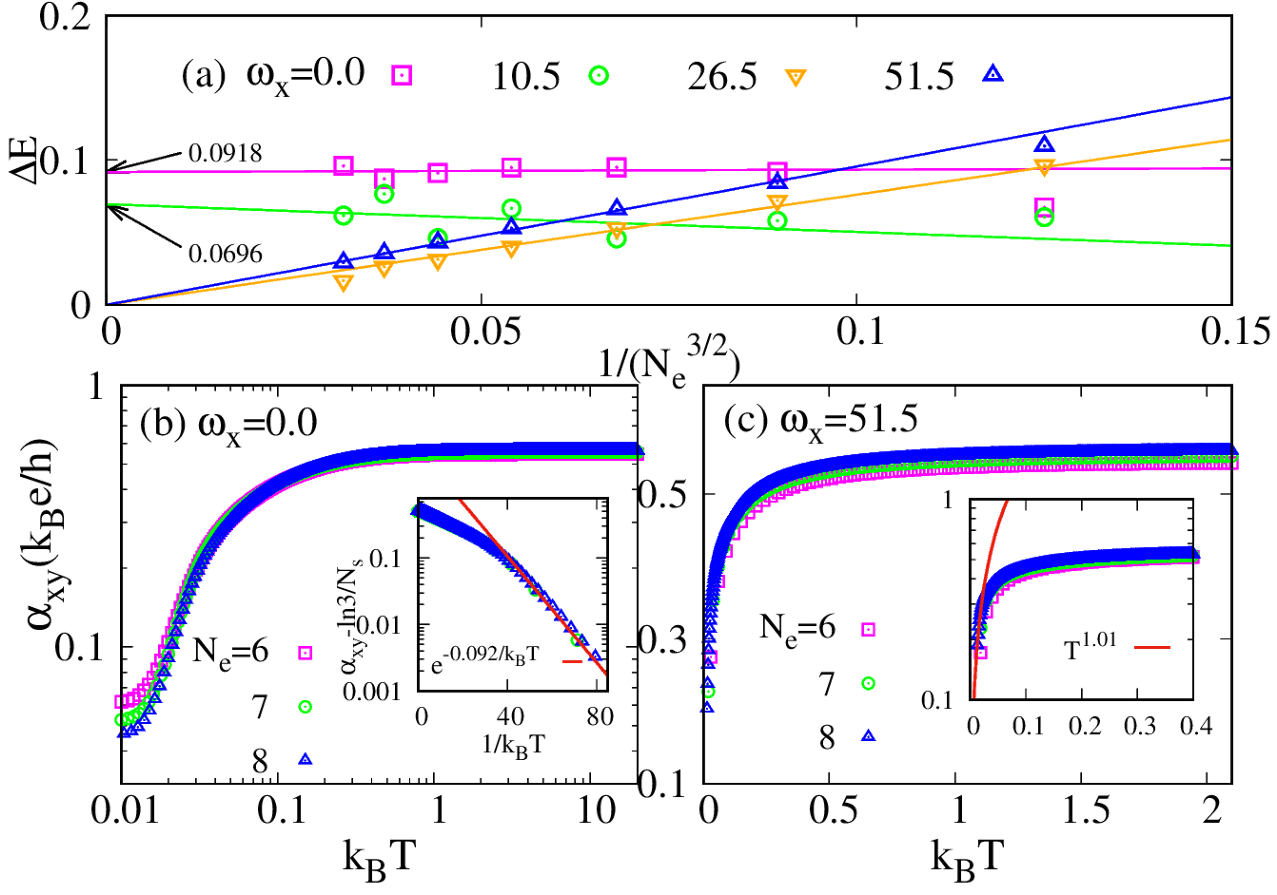}
\caption{\label{gap} (a) Finite-size scalings of the $\nu=7/3$ ground energy gap at parameters in four phases. (b) and (c) display the thermoelectric Hall conductivity $\alpha_{xy}$ for $6-8$ electrons. In the inserted plot in panel (b), we fit the low temperature data with $\alpha_{xy}\propto  \exp(-\Delta E/k_BT)$. The fitting parameter $\Delta E=0.092$ is consistent with the spectrum gap in Figure (a). In the insets of (c), since the system is gapless, we fit the date by power-law scaling function  $\alpha_{xy}\propto T^\eta$ and obtain $\eta \approx1.01 $. }
\end{figure}

The thermoelectric effects, which allow for the direct transformation of heat energy into electrical power, are both fascinating and useful. In a thermoelectric experiment, a temperature gradient $\triangledown T$ is established, causing the system to generate an electrical current $I$ to counteract its effect. The relationship between them is described by $I_i = -\alpha_{ij} \triangledown_j T$, where $\alpha_{ij}$ denotes the thermoelectric conductivity. In practical trials, it is typical to determine the thermopower $S_{xx}$ and Nernst coefficient $S_{xy}$, which are linked to $\alpha_{ij}$ by the equation $S_{ik} = \alpha_{ij}\rho_{jk}$, where $\rho$ signifies resistivity. In the investigation conducted by Sheng and Fu, \cite{Sheng20} the $\alpha_{xy}$ was calculated for the FQH system in torus geometry. They observed a non-Fermi liquid power-law trend where $\alpha_{xy}$ varies as $T^{\eta}$ with $\eta$ approximately equal to 0.5 for composite Fermi-liquid states at filling fractions $\nu = 1/2$ and $\nu = 1/4$. For the Laughlin FQH state at $\nu = 1/3$, it was determined that $\alpha_{xy}$ decreases exponentially as $\alpha_{xy} \sim \exp(-\Delta E/k_BT)$ with a neutral magneto-roton gap as the temperature nears zero. We conducted a comparable analysis on the fast-rotating dipolar fermions in the FQH regime.\cite{fangzy} Here, we analysis the thermoelectric Hall conductivity scaling at the same place in parameter space as that in Fig.~\ref{spectrum}.  Figure~\ref{gap}(a) illustrates the variation in the size of the gap $\Delta E$ across various system sizes. The gap for each system is extrapolated from the scaling behavior of $\alpha_{xy}$ as varying the temperature as shown in Figure~\ref{gap}(b). Therefore, the thermodynamic limit extrapolation indicates that the A/B phases exhibit a gap, whereas the C/D phases do not. Furthermore, the gapless phase D demonstrates Fermi liquid properties with $\alpha_{xy} \propto T^{1.01}$ and $\eta \sim 1$, suggesting the presence of a new phase in that region.  In Fig.~\ref{phase}(d), the D phase is characterized by two properties. As shown in Fig.~\ref{phase}(b) and Fig.~\ref{overlap}, both the Ising nematic order parameter and wave function overlap are zero in D phase. We will demonstrate that the D phase represents a state of charge density wave (CDW).

\begin{figure}[ht]
 \includegraphics[width=0.48\textwidth]{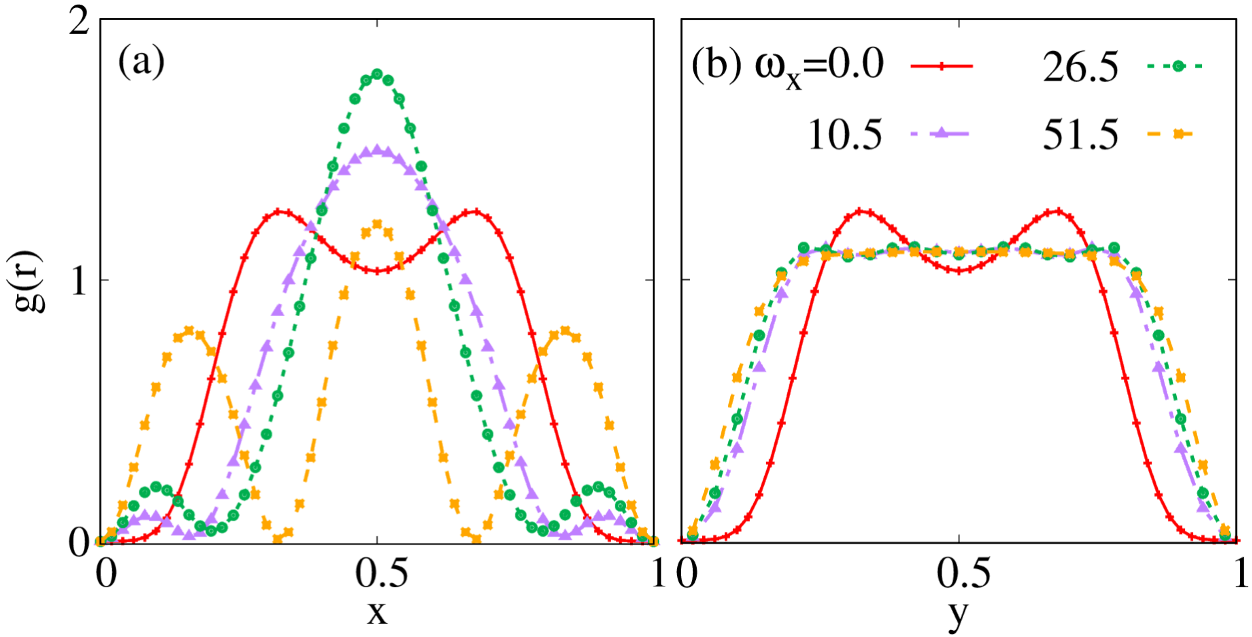}
\caption{\label{gr2d}The ground state pair correlation function $g({\bf r})$ along the (a) $x$ and (b) $y$ directions for four different $\omega_x$s as mentioned previously.  }
\end{figure}

To acquire additional details regarding the four phases, we calculate the pair correlation function as defined by 
\begin{equation}
    g({\bf r})=\frac{L_x L_y}{N_e(N_e-1)}\sum_{i\neq j}\left \langle \psi \right | \delta ({\bf r}-{\bf r_i}+ {\bf r_j})\left | \psi  \right \rangle
\end{equation}
In Fig.~\ref{gr2d}, the $g({\bf r})$ values are plotted along the $x$ and $y$ directions. In phase A, the pair correlation functions display symmetry for $\omega_x=0.0$, indicating rotational symmetry. The lack of symmetry becomes more noticeable in phases B, C, and D. In particular, the values of $g({\bf r})$ along the direction of $y$ are almost identical in the anisotropic phases. In phase D, with $\omega_x = 51.5$, the pair correlation function along the $x$ direction shows a nearly periodic oscillation pattern, a characteristic feature of the CDW state.

\begin{figure}[ht]
 \includegraphics[width=0.5\textwidth]{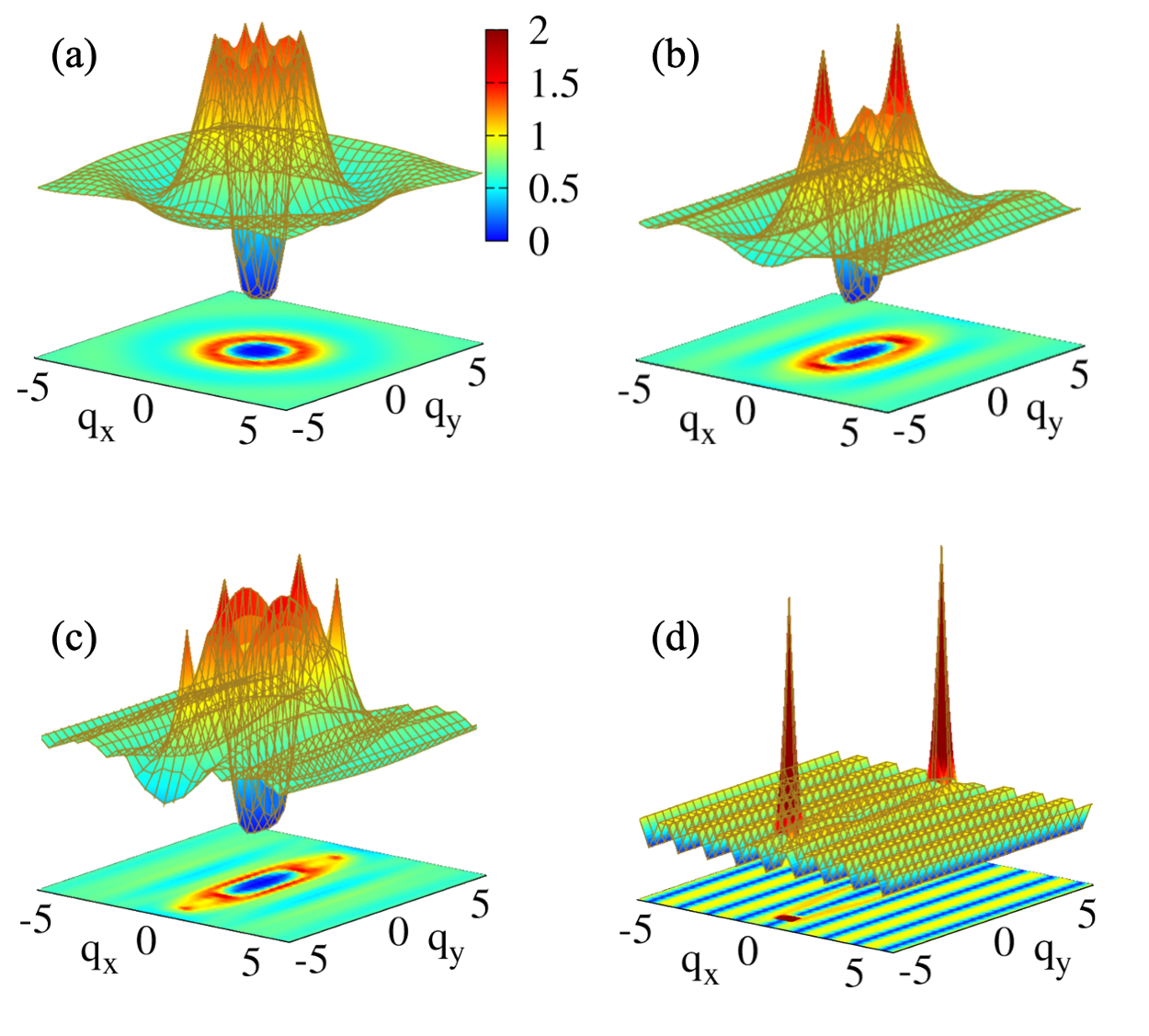}
\caption{\label{sq} The static structure factor for the $\nu= 7/3$ state with $w_0=1.25$ and varying $\omega_x$ values as indicated earlier.}
\end{figure}

As an extra diagnostic tool for the phases of state, it is useful to consider a guiding center structure factor, which is actually the Fourier transformation of the pair correlation function.
\begin{equation}
S_0({\bf q})=\frac{1}{N_e}\sum_{i,j}  \langle e^{i {\bf q}\cdot {\bf r_i}} e^{-i {\bf q}\cdot {\bf r_j}} \rangle  
\end{equation}
The results are shown in Fig.~\ref{sq} also for the same parameters as above in each phase. The transition from a circular to an elliptical contour of $S_0(\bf{q})$ also indicates the breakdown of rotational symmetry. Two sharp peaks appear in rest phases, and they become more and more prominent as increasing the strength of the in-plane field. This is similar to those previously identified in $n \geq 2$ LL states,~\cite{Rezayi99} which are the hallmark of CDW order. In particular, the phase D depicted in Fig.~\ref{sq}(d), the two peaks are divergent and the remaining portion of $S_0(\bf{q})$ are suppressed to almost zero, suggesting the system has transitioned into a fully one-dimensional crystalline state. Those $q$ value of peaks gives the length scale of the unit cell in CDW phase.

\begin{figure}
 \includegraphics[width=0.48\textwidth]{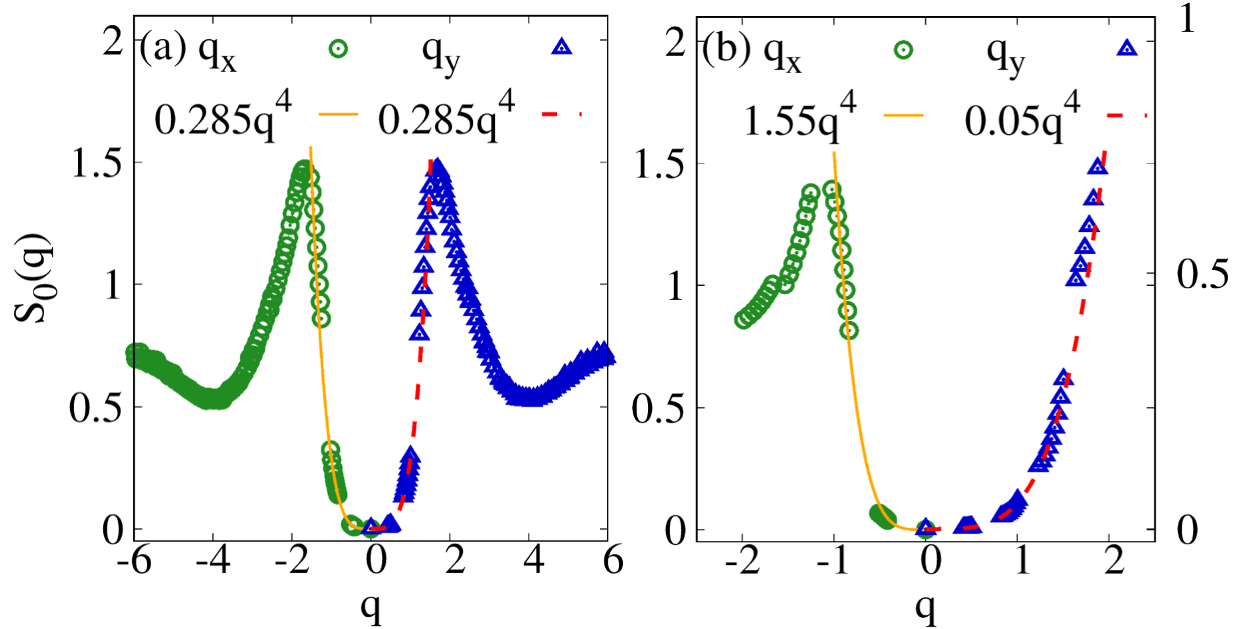}
\caption{\label{sq2} The value of the static structure factor along $q_x$ and $q_y$ direction in Fig.~\ref{sq} (a) and (b) respectively.  }
\end{figure}

To show the presence of the gapped state in both the A and B phases once more, we plot the value of the structure factor along two directions and analyze the behavior near $q = 0$. As shown in Fig.~\ref{sq2}, in phase A, it is expected that the $S_0(\bf{q})$ function exhibits symmetry in two directions, and a behavior proportional to $q^4$ near the point $q = 0$ indicates the presence of an incompressible state with a gap. In phase B, despite the distinct appearance of $S_0(\bf{q})$ in two orientations, its behavior close to $S_0(\bf{q})$ is also quartic, suggesting the presence of an anisotropic FQH state with incompressible properties.

\begin{figure}
 \includegraphics[width=0.48\textwidth]{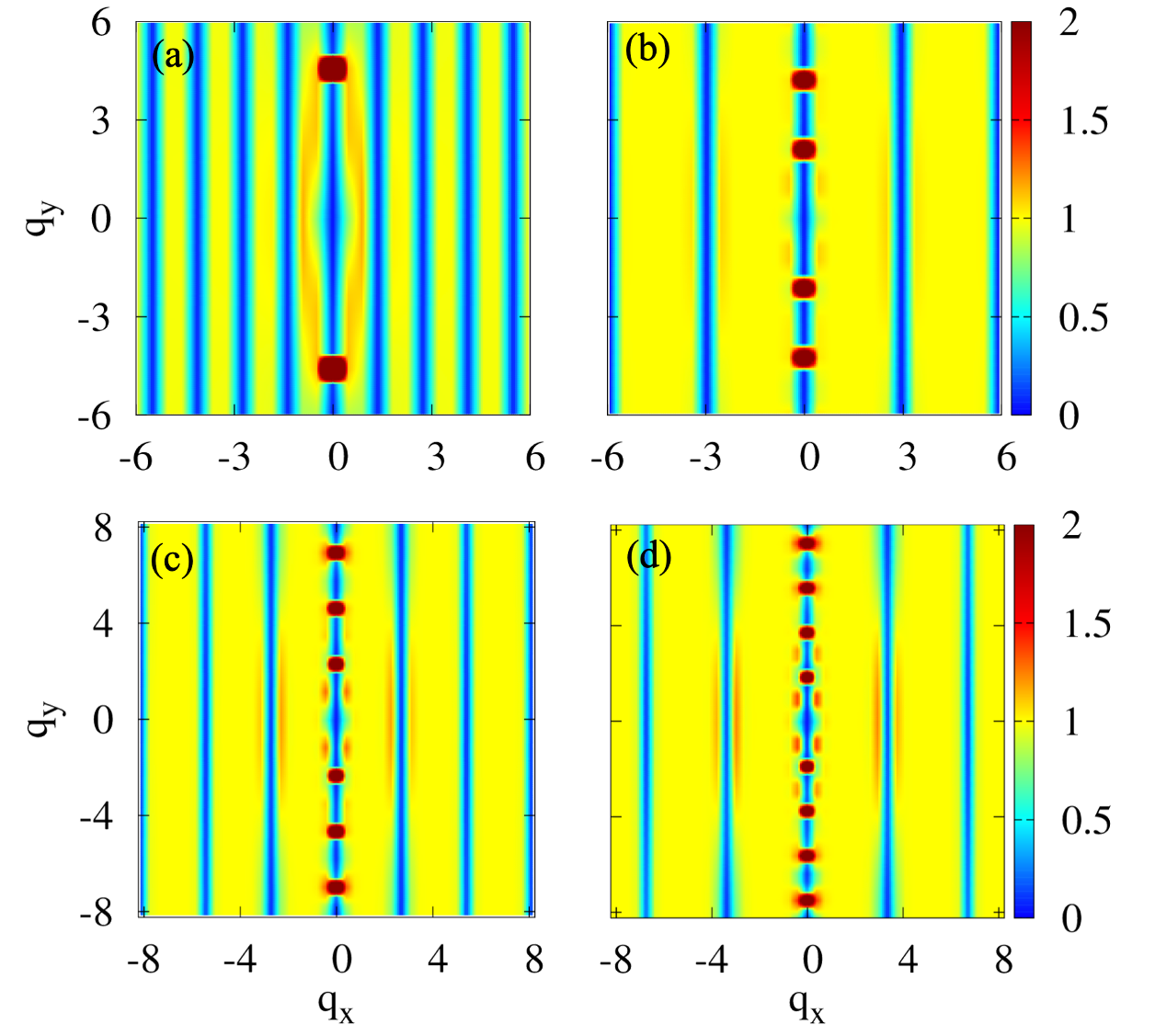}
\caption{\label{fig9} The 2D plot of the static structure factor for the Laughlin filling $\nu= \frac{1}{2n+1}$ $(n=1,2,3,4)$ on the 1LL in parameterized in D phase. }
\end{figure}

To learn more about the CDW phase, we plot the 2D projection of $S_0(\bf{q})$ for several Laughlin fillings $\nu = 1/(2n+1)$ with a large in-plane magnetic field. As illustrated in Fig.~\ref{fig9}, it is interesting to observe that there are $2n$ peaks evenly distributed along a line centered at $q = 0$. This is strongly suggestive of a tendency toward CDW ordering, as a CDW responds strongly to an external potential modulation with a wave vector that matches one of its reciprocal lattice vectors. The fact that
other peaks appear only at integer multiples of the primary wave vector suggests that the CDW has a 1D structure. This is quit nature since we apply the in-plane in one direction. Similar phenomena were also previously observed in Landau levels $n \geq 2$ at half filling.~\cite{Rezayi99}

 \section{Summaries and Conclusions}
In this work, we investigate the fate of the FQH states in a realistic model with considering the effects of the layer thickness and an in-plane magnetic field. Based on solving the single particle problem in this case, the electron-eletron interaction could be expressed in generalized pseudopotentials. After comparing several FQH states in the 1LL, we find that the Ising nematic order reaches its peak at $\nu = 7/3$.  By adjusting the parameters $w_0$ and $\omega_x$, which represent the layer thickness and in-plane magnetic field, respectively, we calculate the neutral gap, Ising nematic order, wave function overlap, pair correlation function, projected static structure factor and finally obtain the phase diagram in the parameter space. As increasing the strength of the in-plane magnetic field, we identify the (A) isotropic FQH, (B) anisotropic FQH, (C) FQH nematic, and (D) CDW phases at $\nu = 7/3$. The spectrum gap and wave function overlap are maxmized at finite $\omega_x$, indicating that small in-plane magnetic field enhances the stability of the Laughlin state at $\nu = 7/3$.

In the FQHN phase, we show through numerical analysis that the rotational symmetry is broken due to a non-zero Ising nematic order, and there is an energy gap closing in the thermodynamic limit, as evidenced by examining the scaling properties of the thermoelectric Hall conductivity. The CDW phase is charactered by Fermi liquid behavior of thermoelectric Hall conductivity, periodic oscillation pattern of the pair correlation function and sharp peaks in static structure factor. These findings align with recent resonant inelastic light scattering experimental observations and provide valuable insights into the underlying physics of the FQH states in microscopic Hamiltonians. During the manuscript preparation, we came across a recent study by Ref.~\onlinecite{Songyang} that introduced a microscopic model for FQHN. This model demonstrates the FQH-FQHN-CDW transition by adjusting the shortest-range pseudopotential ($c_0$ for bosons and $c_1$ for fermions) of the Coulomb interaction in the LLL. We argue that transitioning to the 1LL represents a natural approach to reduce the strength of the shortest-range pseudopotential when compared to the LLL. In our approach, the phase of the fractional quantum Hall nematics is induced by an in-plane magnetic field rather than occurring spontaneously. Our discoveries could provide insights for similar experiments, particularly those involving tilted fields.
 
\acknowledgments
 This work was supported by National Natural Science Foundation of China Grant No. 11974064 and 12347101, the Chongqing Research Program of Basic Research and Frontier Technology Grant No. cstc2021jcyjmsxmX0081, Chongqing Talents: Exceptional Young Talents Project No. cstc2021ycjh-bgzxm0147.

\end{document}